# Extraction of Uncorrelated Sparse Sources from Signal Mixtures using a Clustering Method


Malcolm Woolfson

*Department of Electrical and Electronic Engineering, Faculty of Engineering, University of Nottingham, Nottingham.  NG7 2RD. England*

*email: malcolm.woolfson@nottingham.ac.uk*


**Abstract**


A blind source separation method is described to extract sources from data mixtures where the underlying sources are assumed to be sparse and uncorrelated. The approach used is to detect and analyse segments of time where one source exists on its own. Information from these segments is combined to counteract the effects of noise and small random correlations between the sources that would occur in practice. This combined information can then be used to estimate the sources one at a time using a deflationary method. Probability density functions are not assumed for any of the sources. A comparison is made between the proposed method, the Minimum Heading Change method,  Fast-ICA and Clusterwise PCA. It is shown, for the dataset used in this paper, that the proposed method has the best performance for clean signals if the input parameters are chosen correctly. However the performance of this method can be very sensitive to these input parameters and can also be more sensitive to noise than the Fast-ICA and Clusterwise methods.


# 1        Introduction

One general problem in signal processing is the extraction of individual source signals $\{s_j[n]\}$ from measurements $\{z_i[n]\}$ that are a linear combination of these sources:

$$z_i[n] = \sum_{j=1}^{N} A_{ij} s_j[n] \qquad (1)$$

$$(i = 1, 2, \ldots, M)$$

where $\{A_{ij}\}$ are the mixing coefficients, $M$ is the number of sets of measurement data and there are $N$ underlying sources. In the case where both the sources and mixing coefficients are unknown, this problem comes under the heading of Blind Source Separation (BSS). There are many applications in this area, for example the analysis of EPR data [1], NMR data [2], fetal ECG monitoring [3] and gene mapping [4].

BSS is an undetermined problem, even when $M \geq N$, as both $\{A_{ij}\}$ and $\{s_j[n]\}$ in Equation (1) are unknown, which means that linear estimation methods cannot be applied. There are various approaches to estimating the sources, to take a few examples: Principal Component Analysis (PCA) [5], forcing higher order cross-cumulants to zero [6] and Independent Component Analysis [7], [8]. In various approaches, the data are normally whitened first, using for example PCA or Gram-Schmidt, and then each transformed data component is normalised to have unity root mean squared (rms) value. The relation between the whitened components $\{e_i[n]\}$ and underlying sources $\{s_i[n]\}$ can be written as

$$e_i[n] = \sum_{j=1}^{N} B_{ij} s_j[n] \qquad (2)$$

where $\{e_i[n]\}, i = 1, 2, \ldots, M$ are the whitened components and $\{B_{ij}\}$ are the mixing coefficients. Whitening makes it easier to separate out the individual components which are assumed uncorrelated. From now on we assume that the sources are uncorrelated and that the data has



been whitened and normalised to unit rms value- the method of whitening we will use is the Gram-Schmidt method.

The approaches to BSS described in [5-8] do not make any assumptions as to how the underlying sources vary with time. However, in many applications, for example monitoring of fetal ECG from multilead measurements, the underlying sources are significant only for a segment of time- such sources are termed "sparse". A looser definition of sparsity is that each source should be dominant over the others for a short period of time. The methods described in the previous paragraph can be applied to the case where the underlying sources are sparse; however, a group of BSS methods have been developed which make use of the sparsity of the sources to extract them – such methods come under the heading of Sparse Component Analysis (SCA) [9-17].

Some methods for SCA make use of the following geometrical interpretation of sparsity. Suppose that we are processing two mixtures of two sources, so that $M = N = 2$ in (1). If one of the sources exists on its own for a segment of time then, if we plot one set of data against the other, the resulting phase plot will be a straight line during the time that that source is sparse.

Let us look at the simplest case of two mixtures of two sources that are non-overlapping in time.

Each source is modelled as a Gaussian truncated in time:

$$s_i(t) = a_i \exp\left[\frac{-(t - t_{0i})^2}{2\sigma_i^2}\right] \quad \text{for } |t| \leq 4\sigma_i \tag{3}$$

$$= 0 \quad \text{for } |t| > 4\sigma_i$$

The following parameters are used for each source:

Source 1: $a_1 = 1$, $t_{01} = 0.1$ s, $\sigma_1 = 12.5$ ms



Source 2: $a_2 = 0.1$, $t_{01} = 0.026$ s, $\sigma_1 = 6.25$ ms

The simulated sources are shown in Figure 1(a); in this case the two sources are completely sparse.

These sources are mixed where the randomly chosen mixing coefficients $\{A_{ij}\}$ in Equation (1) are given by the matrix:

$$A = \begin{pmatrix} 1.3 & 2 \\ 1 & 2.85 \end{pmatrix} \tag{4}$$

A sampling frequency of 250 Hz is used.

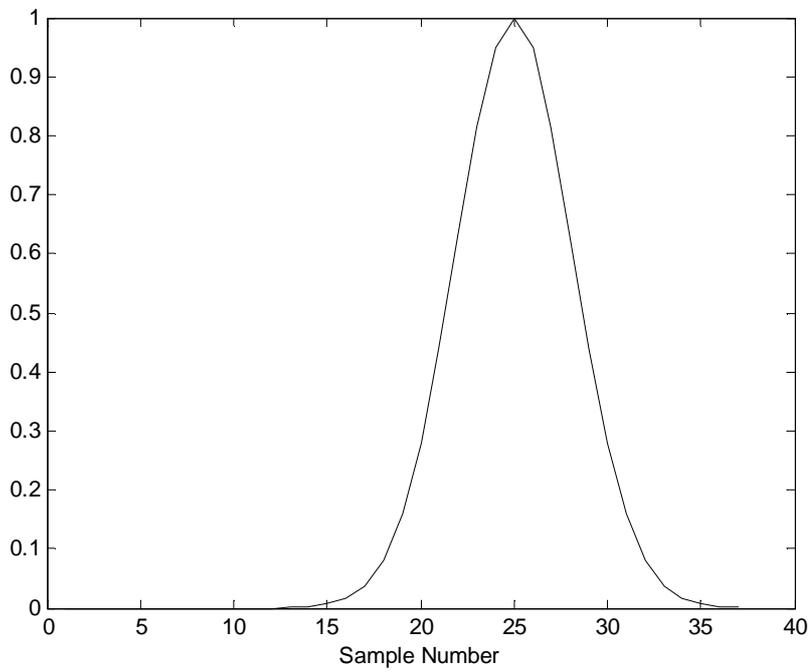



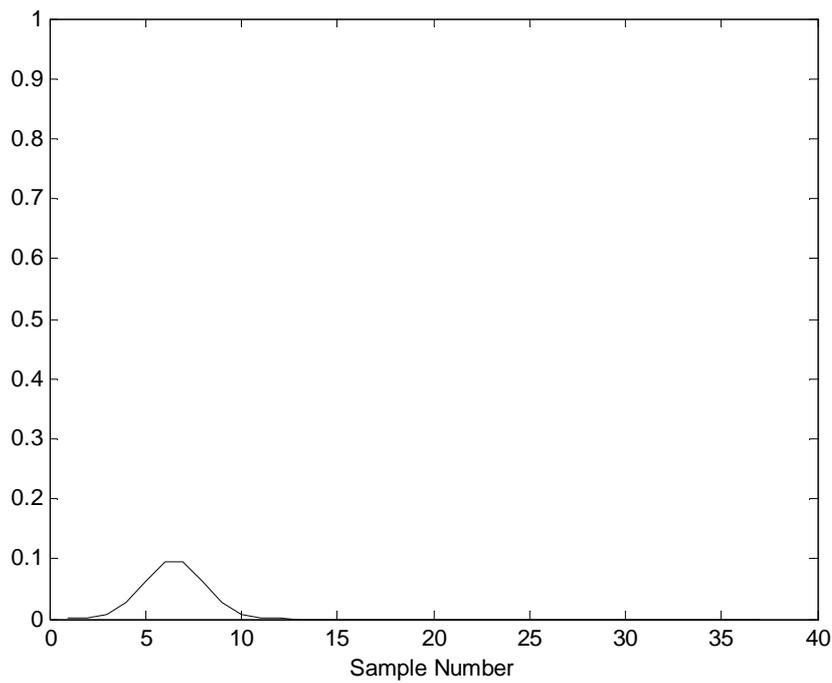

*Figure 1(a)    Simulated Sparse Sources: top figure Source 1, lower figure Source 2*

If one whitens the data using the Gram-Schmidt method, and plots the mixed signal $e_1$ against $e_2$ (Equation (2)) then the phase plot in Figure 1(b) is obtained. It can be seen that the points cluster in two directions, each direction corresponding to a particular source and the principal directions are orthogonal because the underlying sources are uncorrelated.



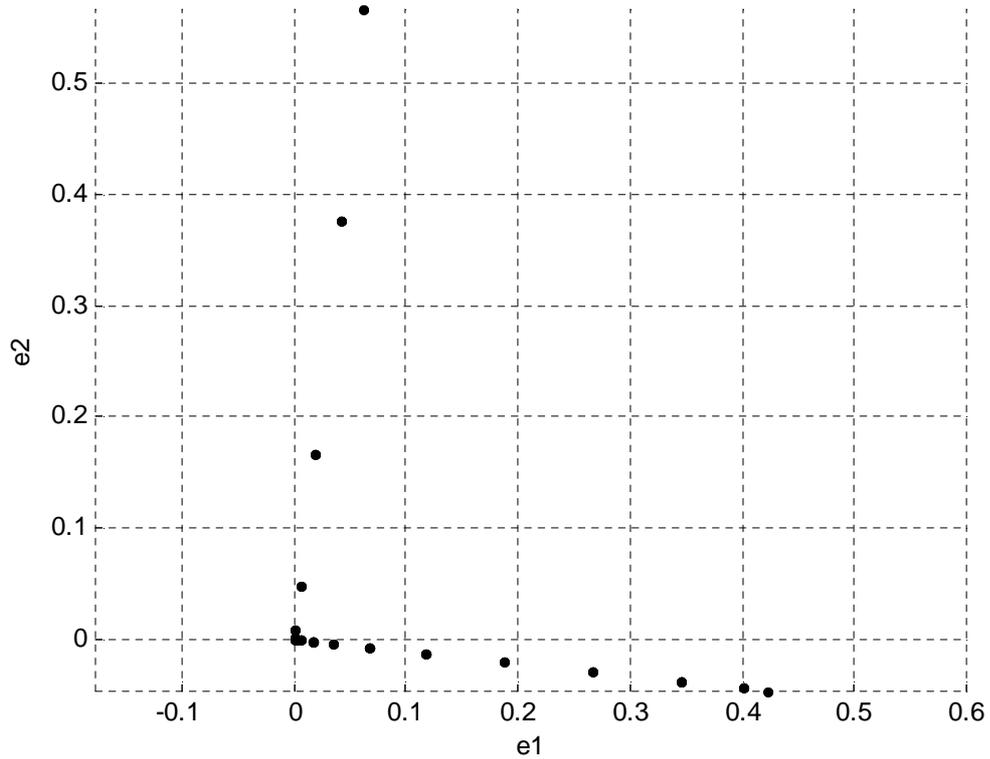

*Figure 1(b) – Plot of e₂ against e₁ after whitening and normalisation*

The principal directions of this plot are directly related to the coefficients $\{B_{ij}\}$ in (2). Several clustering methods have been developed to detect those segments of the plot where straight lines occur, and these directions can then be used to estimate the underlying sources [9],[12]. This method works best when all sources can be detected in the original data. However, sometimes certain sparse sources can be masked by one or more other sources and this will result in errors in the estimation of the underlying sources. Another approach [15,16], adopted in this paper, is to estimate one at a time, using deflation, the vectors in the phase plot corresponding to each sparse source. Taking the simple case of two sources and two data mixtures, suppose that for the example in Figure 1(b), when $e_2[n]$ is plotted against $e_1[n]$, that points in the phase plot are joined up with straight lines in order of time as shown in Figure 2.



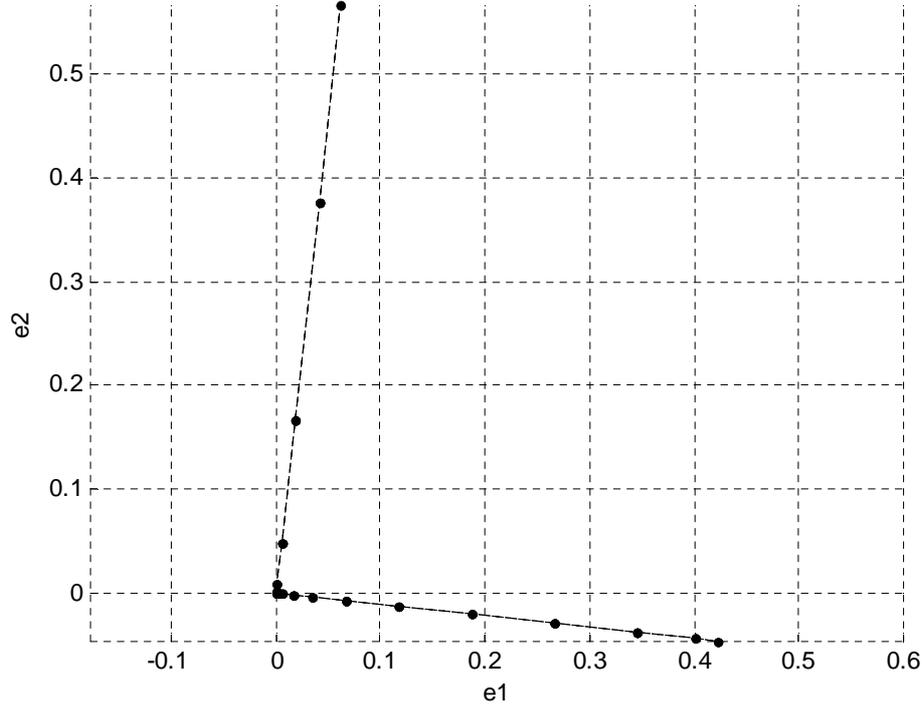

*Figure 2 – Phase plot with points joined up*

Let the vector $\mathbf{e}[n]$ be defined as:

$$\mathbf{e}[n] = (e_1[n], e_2[n]) \tag{5}$$

The "velocity vectors" for the phase plot are defined as

$$\mathbf{v}[n] = \mathbf{e}[n] - \mathbf{e}[n-1] \tag{6}$$

with the normalised heading vector given by

$$\hat{\mathbf{r}}[n] = \frac{\mathbf{v}[n]}{\left|\mathbf{v}[n]\right|} \tag{7}$$

In [15], segments corresponding to a particular source are recognised by looking for three consecutive points $M$ - 2, $M$ - 1 and $M$ where the magnitude of the change in normalised "heading" is a minimum over all the data points; this is deemed to correspond to a sparse source and the estimate of the heading is taken as the most recent one found: $\hat{\mathbf{r}}[M]$. We then estimate the direction in the phase plot corresponding to source 1, $\hat{\mathbf{R}}_1$, from

$$\hat{\mathbf{R}}_1 = \hat{\mathbf{r}}[M] \tag{8}$$



The estimate of source 1 is given by

$$\tilde{s}_1[n] = \hat{\mathbf{R}}_1 . \mathbf{e}[n] \tag{9}$$

Assuming uncorrelated sources, the directions in the phase plot corresponding to the other sparse sources are orthogonal to $\hat{\mathbf{R}}_1$ and hence $\tilde{s}_1[n]$ will pick up contributions from source 1 only and not the other sources. This estimate of source 1 can be subtracted from the phase plot as follows:

$$\mathbf{z}'[n] = \mathbf{e}[n] - \tilde{s}_1[n]\hat{\mathbf{R}}_1 \tag{10}$$

The algorithm is then applied to $\{\mathbf{z}'[n]\}$ to estimate source 2. This method, called the Minimum Heading Change (MHC) method, can be extended to mixtures of $N$ sources; a full description of the method is contained in [15].

This method was successfully applied to mixtures of uncorrelated sparse sources [15] and was extended to correlated sparse sources in [16]. However, as discussed in these two references, the MHC method can be more sensitive to noise than other methods. To illustrate the problem, in Figure 3 the phase space plot of the whitened components corresponding to Figure 2 is shown where noise is added on to $z_1$ and $z_2$ in Equation (1).



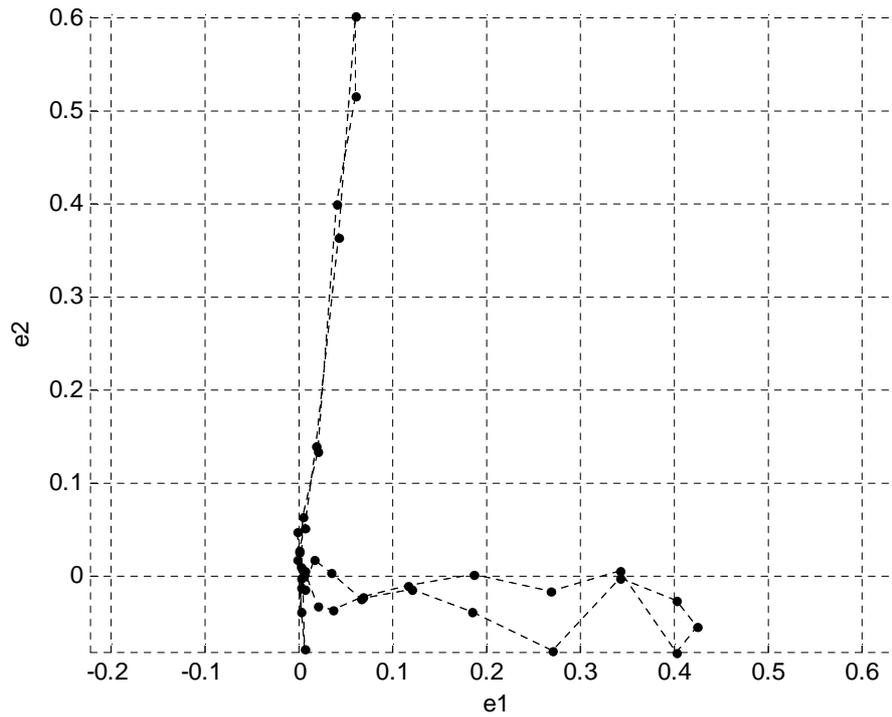

*Figure 3 – Phase plot when noise is added*

The directions in the phase plot corresponding to the two sources are still clear but it can now be seen that noise is affecting the plots. This noise will cause a problem for the MHC in detecting these directions. Only one heading value is chosen in this direction which may deviate significantly from the actual heading. The question to be asked is whether one can obtain more robustness to noise if one clusters <u>all</u> heading vectors over the whole phase plot corresponding to each dominant source direction, similar to the clustering of points used in [12]. One could then perform a weighted average over the heading vectors to obtain a smoother estimate of the underlying sources. We will refer to this approach as the Global Method.

It should be pointed out that although the deflation approach has the advantage of being able to estimate sparse sources that are hidden in the original data, the required iterative procedure means that estimation errors can accumulate when separating more and more sources.



The aim of this paper is to derive a method based on clustering to find the dominant directions in the phase plot and to assess any improvements of this method compared to the MHC derived in [15] when there is noise present for data consisting of mixtures of uncorrelated sources.

## 2    Motivation of the Clustering Algorithm

In the MHC method, we are looking at the normalised headings in phase space

$$\hat{\mathbf{r}}[n] = \frac{\mathbf{v}[n]}{|\mathbf{v}[n]|} \tag{11}$$

where

$$\mathbf{v}[n] = \mathbf{e}[n] - \mathbf{e}[n-1] \tag{12}$$

For sources which are sparse, we wish to cluster together headings corresponding to each source; this means that we are clustering headings which are deemed to be "close" enough to each other. The headings that are clustered may come from different non-adjacent segments of the data.

We now need to consider how to cluster the headings corresponding to each source. Conventionally, clustering methods are applied to the original data and various iterative approaches are used, such as K-means clustering. However, in this paper, a much simpler approach can be used: as the differences between adjacent data points are being analysed, then one can directly associate similar heading vectors across the whole data. One way is to compute the magnitude of the differences between each pair of headings and look for close associations between these pairs. However, this approach has been found to be computationally expensive. An alternative method, which is adopted in this paper, is to cluster the headings component by component using a sorting method.



In the clustering procedure that is adopted in this paper, the following initial steps are carried out. The normalised heading vector at sample point $n$ can be written in terms of its component values as:

$$\hat{\mathbf{r}}[n] = (\hat{r}_1[n], \hat{r}_2[n], ..., \hat{r}_N[n]) \qquad (13)$$

Suppose that we calculate the magnitude of each heading component and then sort each heading component in ascending order of magnitude:

$$\{\hat{r}^{sort}{}_i[m]\} = \text{sort}\{|\hat{r}_i[n]|\} \qquad i=1,2,...,N \qquad (14)$$

Next we plot each sorted normalised heading component as a function of heading index, $m$, in the reordered sequence; the resulting plot will be different depending on the sparsity of the sources. To see this, let us look at three examples.

In the first two examples source $s_1$ and $s_2$ are generated by a uniform random number generator over 100 samples.

*(i) Both data inputs are identical to within a scaling constant*

In the first example, suppose that the two data inputs are equal to within a scaling constant so that there is perfect correlation between the two inputs. The resulting ordered plots for the case where the scaling constant is 2 for $\{\hat{r}^{sort}{}_i[m]\}$ are shown, for each component in Figure 4(a).



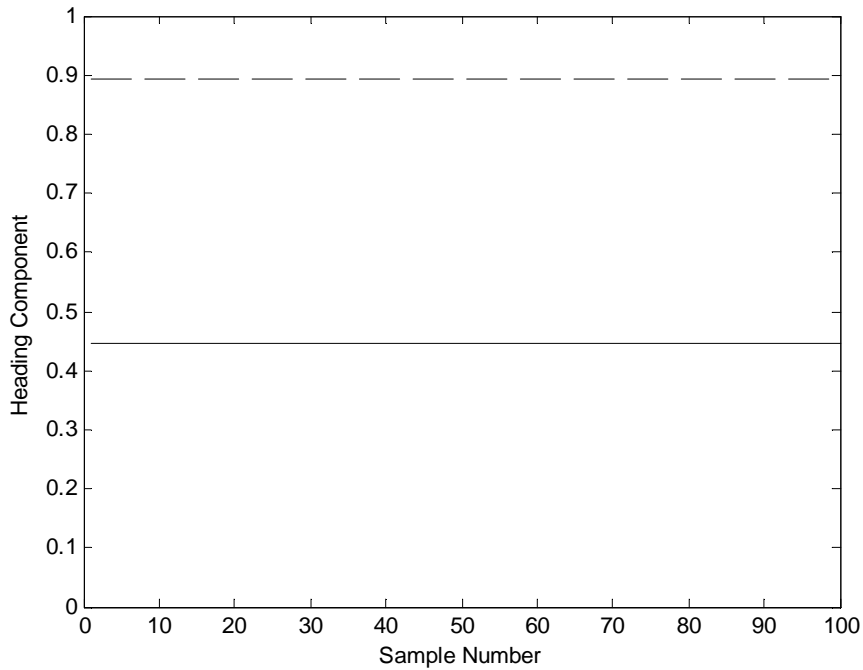

.

*Figure 4(a)- Both inputs same sequence of random numbers.*
*Full line: Component 1, Dashed Line: Component 2.*

The normalised heading components will then be identical for all data points. As expected, the sorted normalised heading components form a horizontal line in these plots indicating that the two inputs are perfectly correlated.

*(ii)     Both data inputs are uncorrelated*

In the second example, one data mixture is equal to source 1 (used in the previous example) and the other equal to source 2. In this case example, the headings components are uncorrelated between components, so that if we plot the heading components in ascending order they are monotonically increasing as shown in Figure 4(b).



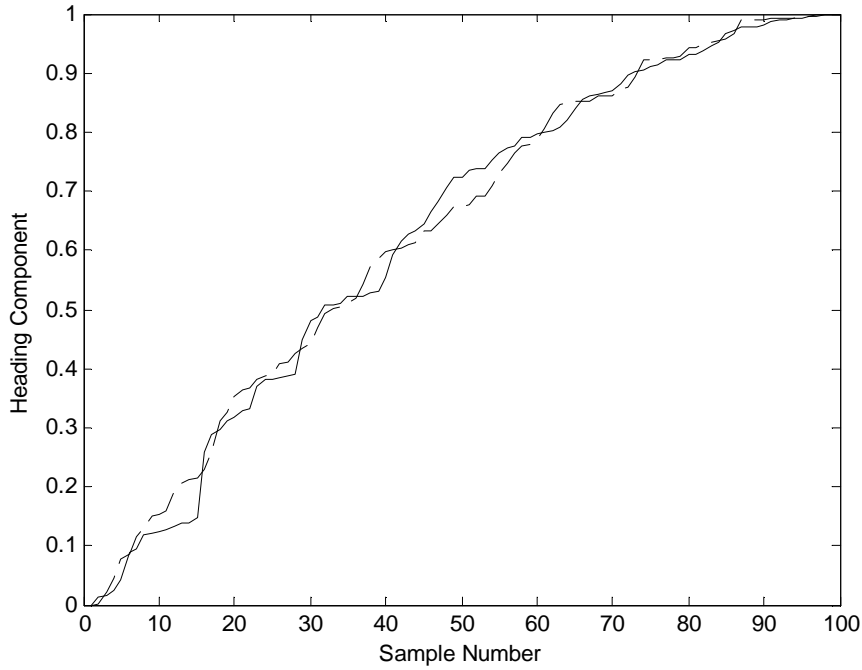

*Figure 4(b) – Inputs are different sequences of random numbers*
*Full line: Component 1, Dashed Line: Component 2.*

*(iii)    Data inputs are mixtures of sparse sources*

For the sake of example, let the two sources be $\{u_1[n]\}$ and $\{u_2[n]\}$, which are each generated by a uniform random number generator.   Source 2 is shifted to the right by 90 samples, with 90 zero values added at the beginning so that the shifted data is given by

$$u_2'[n] = u_2[n-L] \quad \text{for } L < n \le 100 + L$$
$$u_2'[n] = 0 \qquad\qquad \text{for } n \le L \tag{15}$$

where $L = 90$.

Zeros are added to the end of Source 1 as follows:

$$u_1'[n] = 0 \qquad\qquad \text{for } 100 < n \le 100 + L$$
$$u_1'[n] = u_1[n] \qquad \text{for } 1 \le n \le 100 \tag{16}$$

Hence the two sources are now sparse.

 The two sources are mixed using the randomly selected mixing matrix:



$$A = \begin{pmatrix} 0.799 & -0.498 \\ -0.373 & -0.133 \end{pmatrix} \qquad (17)$$

The resulting phase plot of the data, after whitening, is shown in Figure 4(c) where it can be seen that there is a mixture of straight line segments, indicated by "1" and "2" and a random pattern elsewhere.

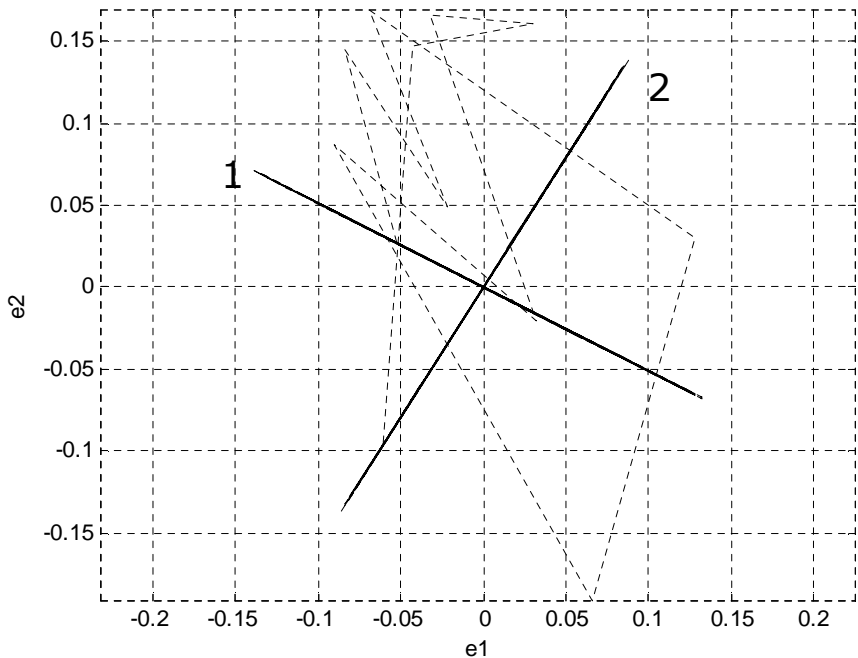

*Figure 4(c) – Gram-Schmidt plot of two sources*

The resulting ordered plots for $\{\hat{r}^{sort}_i[m]\}$ are shown in Figure 4(d).



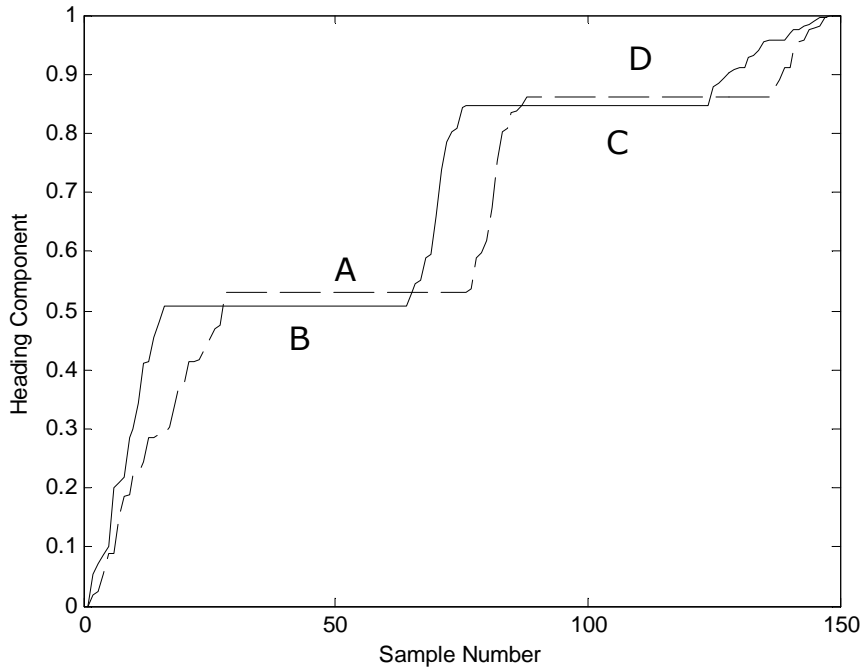

*Figure 4(d) - Mixture of Sparse Sources*
*Full line: Component 1, Dashed Line: Component 2.*

In this third case, we can see that the resultant ordered normalised heading components

consist of monotonically increasing values characteristic of randomness along with horizontal

segments reflecting the data segments 1 and 2 in Figure 4(c) where one source exists on its

own.  It is the heading component values in these horizontal sections that we are interested in.

Now the heading components in each of the segments A and B in Figure 4(d) will come from

two different sources, as will those in C and D.  But the question is which of the headings in

A are associated with C and D?  A similar question can be asked for the headings in segment

B.  As there are only two sources in this example, and the magnitude of the normalised

heading vectors are one, then it can be deduced that segment A is associated with segment C

and B is associated with D.  However, if there are three or more sources, then it would not be

so easy to make this association; to address this problem we need to reorder the heading

components as a function of time and look for sample values where both heading components



are simultaneously associated with one of the horizontal components in Figure 4(d). The heading vectors associated with each source can then be averaged over in some way to reduce the effects of noise. The potential advantage of this clustering method over the local MHC method is that, in the former method, we can associate headings that are not adjacent; in principle estimating the average heading over several headings should be more accurate in the presence of noise compared with just using the more localised MHC method.

## 3      Development of the Clustering Algorithm

The steps of the clustering algorithm are as follows:

(1)      Input whitened data $\{e_i[n]\}$,($i = 1,\ldots,N$)

(2)      For each component $i$ calculate velocity vectors from adjacent data points:

$$\mathbf{v}[n] = \{v_i[n]\} = \{e_i[n] - e_i[n\text{-}1]\}$$

To illustrate the clustering approach used in this work, we will use the following simplified example where there are 10 headings to sort into clusters and the number of sources, $N = 2$, with the velocity components $(v_1[n], v_2[n])$ given in Table 1.

| Heading Number, $n$ | $v_1[n]$ | $v_2[n]$ |
|:---:|:---:|:---:|
| 1 | 1 | 2 |
| 2 | -2 | 3 |
| 3 | 1 | 2 |
| 4 | 2 | 4 |
| 5 | 5 | 3 |
| 6 | -1 | -2 |
| 7 | -4 | 6 |
| 8 | 5 | 5 |
| 9 | -4 | 6 |
| 10 | 5 | 10 |

*Table 1 – Heading Components*

(3)      Calculate heading vector

$$\hat{\mathbf{r}}[n] = \frac{\mathbf{v}[n]}{\left|\mathbf{v}[n]\right|}$$



(4)     Take absolute values of each component of the normalised heading vector: $|\hat{r}_i[n]|$

After carrying out Steps (3) and (4) we obtain the magnitude of the heading

components as shown in Table 2.

| Normalised Heading Number, $n$ | $|\hat{r}_1[n]|$ | $|\hat{r}_2[n]|$ |
|---|---|---|
| 1 | 0.4472 | 0.8944 |
| 2 | 0.5547 | 0.8321 |
| 3 | 0.4472 | 0.8944 |
| 4 | 0.4472 | 0.8944 |
| 5 | 0.8575 | 0.5145 |
| 6 | 0.4472 | 0.8944 |
| 7 | 0.5547 | 0.8321 |
| 8 | 0.7071 | 0.7071 |
| 9 | 0.5547 | 0.8321 |
| 10 | 0.4472 | 0.8944 |

*Table 2 – Magnitude of Normalised Heading Components*

(5)     For each component $i$, sort absolute values of normalised heading components in

ascending order

$\{\hat{r}^{sort}_i[m]\} = \text{sort}\{|\hat{r}_i[n]|\}$

Note the following mapping between $n$ and $m$ for each component $i$

$f_{1i}[m] = n$

The sorted headings and the array $\{f_{1i}[m]\}$ for the above diagram are shown in

Table3:

| Sorted Normalised Heading Number, $m$ | $|\hat{r}^{sort}_1[m]|$ | $f_{11}[m]$ | $|\hat{r}^{sort}_2[m]|$ | $f_{12}[m]$ |
|---|---|---|---|---|
| 1 | 0.4472 | 1 | 0.5145 | 5 |
| 2 | 0.4472 | 3 | 0.7071 | 8 |
| 3 | 0.4472 | 4 | 0.8321 | 2 |
| 4 | 0.4472 | 6 | 0.8321 | 7 |
| 5 | 0.4472 | 10 | 0.8321 | 9 |
| 6 | 0.5547 | 2 | 0.8944 | 1 |
| 7 | 0.5547 | 7 | 0.8944 | 3 |
| 8 | 0.5547 | 9 | 0.8994 | 4 |
| 9 | 0.7071 | 8 | 0.8944 | 6 |
| 10 | 0.8575 | 5 | 0.8944 | 10 |

*Table 3 – Sorted Heading Components*



(6)  Cluster heading components by looking at the differences between adjacent values of $\hat{r}^{sort}{}_i[m]$ given by $\left|\hat{r}^{sort}{}_i[m] - \hat{r}^{sort}{}_i[m-1]\right|$.

Choose a threshold ε.

If $\left|\hat{r}^{sort}{}_i[m] - \hat{r}^{sort}{}_i[m-1]\right| < \varepsilon$

$\qquad C_i[m] = 1$

$\qquad\qquad$ else

$\qquad C_i[m] = 0$

Let **C** be a matrix for which $C_i[m]$ is the element in the $m^{th}$ row and $i^{th}$ column.

In the above example there is assumed to be no noise. In practice, there will be noise and that is the reason why we allow the difference between adjacent values of $\hat{r}^{sort}{}_i[m]$ to be less than some non-zero threshold ε; we will discuss later how to choose ε.

For our example, the values of $\{C_i[n]\}$ and $f_{1i}$ are shown in Table 4.

| Sorted Normalised Heading Number, $m$ | $C_1[m]$ | $f_{11}[m]$ | $C_2[m]$ | $f_{12}[m]$ |
|---|---|---|---|---|
| 1 | 0 | 1 | 0 | 5 |
| 2 | 1 | 3 | 0 | 8 |
| 3 | 1 | 4 | 0 | 2 |
| 4 | 1 | 6 | 1 | 7 |
| 5 | 1 | 10 | 1 | 9 |
| 6 | 0 | 2 | 0 | 1 |
| 7 | 1 | 7 | 1 | 3 |
| 8 | 1 | 9 | 1 | 4 |
| 9 | 0 | 8 | 1 | 6 |
| 10 | 0 | 5 | 1 | 10 |

*Table 4 – C values*



(7)    Look for column $j$ in $\{C_i[m]\}$ with largest number of adjacent values of 1's:

*Notes*

(i)    In Table 4 it can be seen that there are two such clusters of 1's: $C_1[m]$ for $2 \leq m \leq 5$ and $C_2[m]$ for $7 \leq m \leq 10$. In this case, the software picks up the cluster of $C_1$ values but the same final result will be obtained if the other cluster is picked first.

(ii)    Note that $C_1[1] = C_1[6] = C_2[3] = C_2[6] = 0$. The reasons for putting zeros at these points in the Table is to separate clusters of 1's corresponding to different headings; for example, if $C_1[6]$ was put equal to 1 then there would be a continuous cluster of 1's from $C_1[2]$ to $C_1[8]$ implying that all these components come from the same heading which clearly they do not.

(8)    Now that we have identified a clustering of a heading component in Table 4, we now need to associate these components to the time ordered components shown in Table 2. This is where the values of $f_{11}[m]$ and $f_{12}[m]$ in Table 4 are used. In this Table, the following normalised sorted heading values are clustered together $|\hat{r}^{sort}_1[2]|$,

$|\hat{r}^{sort}_1[3]|$, $|\hat{r}^{sort}_1[4]|$ and $|\hat{r}^{sort}_1[5]|$. Following on from Note (ii) in Step (7) above we should also include $|\hat{r}^{sort}_1[1]|$ in the clustering. Using the values of $f_{11}[m]$ in this Table, these sorted heading components correspond to the following unsorted heading components: $|\hat{r}_1[1]|$, $|\hat{r}_1[3]|$, $|\hat{r}_1[4]|$, $|\hat{r}_1[6]|$ and $|\hat{r}_1[10]|$. We now define a matrix $\mathbf{C}^{\mathbf{U}}$, where the element in the $n^{th}$ row and $i^{th}$ column is $C^U_i[n]$; we fill in 1's in column 1 at rows 1,3,4,6 and 10 indicating that the clustering of the first heading component has taken place, as shown in Table 5.



| Normalised Heading Number, $n$ | $C^U{}_1[n]$ | $C^U{}_2[n]$ |
|---|---|---|
| 1 | 1 | |
| 2 | 0 | |
| 3 | 1 | |
| 4 | 1 | |
| 5 | 0 | |
| 6 | 1 | |
| 7 | 0 | |
| 8 | 0 | |
| 9 | 0 | |
| 10 | 1 | |

*Table 5 - $C^U$*

(9)     Now, we have found that the largest cluster for component 1 corresponds to the original sample numbers 1, 3, 4, 6 and 10.  We now need to determine how many heading values at these sample numbers for component 2 are also in a cluster.  To determine this, look at Table 4.  We can see that the values of $C_2[n]$ corresponding to these time points are 0,1,1,1,1.  However, for the reason stated in Step 7(ii), we need to put $C_2[6] = 1$ as this is  part of the same cluster corresponding to $C_2{}^U[1] = 1$ .  Hence the second column of the above table can be filled in as follows:

| Normalised Heading Number, $n$ | $C^U{}_1[n]$ | $C^U{}_2[n]$ |
|---|---|---|
| 1 | 1 | 1 |
| 2 | 0 | 0 |
| 3 | 1 | 1 |
| 4 | 1 | 1 |
| 5 | 0 | 0 |
| 6 | 1 | 1 |
| 7 | 0 | 0 |
| 8 | 0 | 0 |
| 9 | 0 | 0 |
| 10 | 1 | 1 |

*Table 6 – Time ordered values for $C^U$*



(10)   In Table 6, we are looking for rows where all elements $\{C^U_i\}$ are 1; in this case both

components are part of a cluster.  This can be achieved by performing a logical AND

of the elements of each row to produce the following Table:

| Normalised Heading Number, $n$ | $D[p]$ |
|---|---|
| 1 | 1 |
| 2 | 0 |
| 3 | 1 |
| 4 | 1 |
| 5 | 0 |
| 6 | 1 |
| 7 | 0 |
| 8 | 0 |
| 9 | 0 |
| 10 | 1 |

*Table 7 – D[p]*

where  $D[p] = $ AND $(C^U_1[p], C^U_2[p])$

(11)   Each row, $p$, where $D[p] = 1$ corresponds to a heading in the same cluster; for the

above example, the following velocity components form a cluster:

$\mathbf{v}[1]$, $\mathbf{v}[3]$, $\mathbf{v}[4]$, $\mathbf{v}[6]$ and $\mathbf{v}[10]$

which agrees with Table 2.

*Comment*: One could look for other clusters of headings in Table 4, but in this paper

we look at using a deflation approach where we estimate each heading iteratively and

subtract the corresponding estimated source from the data as in Equation (10), the

algorithm is applied to the data $\{\mathbf{z}'[n]\}$ and the clustering method is applied again to

the remaining data.  This method is applied until there are no further sources to

estimate. This subtraction process should make it easier to isolate the other

components.



## 4 Estimation of Heading Vector

Let $\mathbf{v}^a = (v_1{}^a, v_2{}^a, ..., v_N{}^a)$ be the actual non-normalised heading vector (which we also call the velocity vector).

Suppose that the clustering algorithm has been carried out on each component and let us look at velocity component $i$, $v_i^a$. Let us assume that there are $J$ velocity vectors within a cluster. Let $\hat{\mathbf{r}}^{\mathbf{e}} = \{\hat{r}_i{}^e\}$ denote the estimate of the heading from a particular cluster of headings. We now need to estimate $\{\hat{r}_i{}^e\}$ from the set of velocity vectors that have been found from the clustering method: $\{v_i{}'[1]\}, \{v_i{}'[2]\}, ..., \{v_i{}'[J]\}$, where $v_i{}'[n]$ is the $i^{th}$ component of the $n^{th}$ velocity vector.

Each velocity component will be affected by noise. One possibility to determine the $i^{th}$ component of the estimated heading, $\hat{r}_i{}^e$, is to perform a direct average over $j$ of $v_i{}'[j]$. However this is not optimal for the following reason.

The relation between the clustered and actual $i^{th}$ velocity component is given by

$$v_i{}'[j] = v_i^a[j] + n[j] \tag{18}$$

where $j$ refers to this velocity being the $j^{th}$ member of the cluster and $n[j]$ is a sample of noise, assumed Gaussian. The actual velocity to noise ratio is given as

$$VNR[j] = \frac{|v_i^a[j]|}{\sigma} \tag{19}$$

where $\sigma$ is the standard deviation of noise.

The corresponding relation for the $k^{th}$ member of the cluster is:

$$v_i{}'[k] = v_i^a[k] + n[k] \tag{20}$$

with velocity to noise ratio

$$VNR[k] = \frac{|v_i^a[k]|}{\sigma} \tag{21}$$



Because the magnitudes of any two velocities in the cluster may be different, then in general:

$$VNR[j] \neq VNR[k] \qquad (22)$$

Noise will affect smaller magnitude velocities more than those with larger magnitude. Hence any estimator should take this into account by putting more weight on larger heading components than smaller heading components when averaging over all components in a cluster, because the VNR for the larger headings are larger. This problem is addressed in [18] for the averaging of evoked potentials, where it is shown that the estimate, $\tilde{V}_i$, of the $i^{th}$ velocity component is given by

$$\tilde{V}_i = \frac{\sum_{j=1}^{J} M[j].v_i'[j]}{\sum_{j=1}^{J} (M[j])^2}. \qquad (23)$$

where $M[j] = \sqrt{\sum_{i=1}^{N} \{v_i'[j]\}^2}$

When $M[j] = 1$ for all $j$, this reduces to a straight average over all headings.

The above processing is applied to each velocity component $i$ (=1,…,$N$) so that the estimate of the velocity vector becomes:

$$\tilde{\mathbf{V}} = (\tilde{V}_1, \tilde{V}_2, ..., \tilde{V}_N)$$

The estimate of the normalised heading vector is then

$$\hat{\mathbf{R}}_\mathbf{e} = \frac{\tilde{\mathbf{V}}}{|\tilde{\mathbf{V}}|} \qquad (24)$$

$\hat{\mathbf{R}}_\mathbf{e}$ is then used in Equation (8-10) in place of $\hat{\mathbf{R}}_\mathbf{1}$.



## 5      Input Parameters

Let the $N$ components of the velocity vector be written as

$$\mathbf{v}[n] = \left(v_1[n], v_2[n], ..., v_N[n]\right)$$

In order to avoid effects of spurious noise, a velocity vector is accepted at sample point $n$ if

$$V_{\max}[n] \geq v^{th}.v_{\max} \tag{25}$$

where

$$V_{\max}[n] = \max\left\{|v_1[n]|, |v_2[n]|, ..., |v_N[n]|\right\}$$

$0 < v^{th} < 1$ is a chosen threshold

and

$$v_{\max} = \max\left\{|\mathbf{v}[1]|, |\mathbf{v}[2]|, ..., |\mathbf{v}[M]|\right\}$$

is the maximum value of the magnitude of the velocity vector over all $M$ sample points.  This threshold is used in both the MHC and Global methods.

We now need to consider the choice of ε in Step 6 of the general clustering algorithm described in Section 3.  This parameter is used to determine if two heading components are associated with the same source.  Looking at Figure 4(d), for example, it can be seen that in the regions where one source is on its own, the ideal value for ε is zero. In practice, because of the effects of noise and other sources, a non-zero value for ε should be chosen.  Referring to Figure 4(b) where the ordered headings are plotted for random noise, it can be seen that the graph is increasing in a non-linear way; this can be crudely approximated as linear, where the sorted normalised heading is given approximately by

$$\hat{r}^{sort}{}_i[m] \approx \frac{m}{M} \tag{26}$$

where $M$ is the number of headings.

Hence the difference between adjacent ordered normalised headings can be approximated by



$$\hat{r}^{sort}{}_i[m] - \hat{r}^{sort}{}_i[m-1] \approx \frac{1}{M} \tag{27}$$

This implies that for association between two sorted headings that we must choose the parameter ε such that

$$\varepsilon = \alpha \frac{1}{M} \tag{28}$$

where $\alpha \leq 1$.

If $\alpha$ is chosen to be too small, then associations between heading components belonging to the same source will not occur; if too large, then too many false associations will occur between headings that are not from the same source. According to Step 10 in Section 3, a cluster is only declared if an association is found between all components of the heading; hence, randomly associated heading components will tend to AND to zero.

Extensive simulations have been carried out to optimise the parameter $\alpha$ in (28) and it has been found that a good compromise value to use is $\alpha = 1$; this is used in all the simulations and data analysis carried out in this paper.

## 6    Results

In this Section, the performance of the Global Method is compared to the following three methods: (i) MHC [15] (ii) Fast-ICA [19] and (iii) Clusterwise PCA [9]. A selection of three sets of data are used: (i) sources that are purely sparse and uncorrelated (ii) sources that are locally sparse and weakly correlated (iii) experimental abdominal and thoracic ECG data taken from an expectant mother. The aim here is to demonstrate when the proposed method works well compared with standard techniques and to also discuss the limitations of this method. It is particularly of interest to compare the robustness to noise of the proposed technique with the other methods.



Fast-ICA is an implementation of the ICA method; this method is not specifically designed to be applied to sparse signals and can be applied to mixtures of non-sparse sources. It is found for the signals studied in this work that best results are obtained with the Fast-ICA method if one uses the deflation approach and Gaussian non-linearity [19]. The Clusterwise PCA method is specifically designed to be applied to mixtures of sparse sources; clustering takes place using minimum component analysis to find the directions in phase space corresponding to each source [9].

Monte Carlo simulations are carried out to compare the robustness of all the methods to noise. Monte Carlo simulations are also carried out for the Fast ICA and Clusterwise PCA methods when applied to clean signals because the former method randomly initialises the weights and the latter uses a *randperm* function to define the hyperplanes in the search process. The following procedure is adopted in this paper to come up with a figure of merit to quantify the difference between the actual and estimated sources.

Firstly, when comparing actual and estimated sources, to take into account possible scaling of the estimates, $\{\tilde{s}_i\}$, they are normalised so that their rms values are 1. This same normalisation procedure is carried out with the actual sources.

Suppose that the actual sources are given by $\{s_i[m]\}$ where $m$ is a sample index. Suppose also that, at the $q^{th}$ Monte Carlo run, the estimates are $\left\{\tilde{s}_i^{\,q}[m]\right\}$. Now, when applying BSS methods, the estimate may be a scaled version of the actual source but it could also be inverted. To take into account possible inversion of the estimates, we compute the cross-correlation function between the $j^{th}$ estimate and $i^{th}$ actual source using the *corrcoef* function of MATLAB. Supposing this value is $c_{ij}^{\,q}$.



At each Monte Carlo run, define the matrix $[C_{ij}{}^q]$ $(i=1,...,M; j=1,...,M)$ where the $ij^{th}$ element is $c_{ij}{}^q$. An association between each source and estimate is determined by finding the row $r$ and column $s$ of the entry in the matrix $[C_{ij}{}^q]$ with the largest magnitude $\left| c_{rs}{}^q \right|$; the actual source $r$ is then associated with the estimate $s$. The error between source $r$ and estimate $s$ is computed as follows:

$$\varepsilon_s{}^q[n] = s_r[n] - \tilde{s}_s[n] \qquad \text{if } c_{rs}{}^q > 0 \qquad\qquad (29)$$

$$= s_r[n] + \tilde{s}_s[n] \qquad \text{if } c_{rs}{}^q < 0$$

where in the above equations, it is assumed that both $\{s_r[n]\}$ and $\{\tilde{s}_s[n]\}$ have been normalised to unity. The second condition is needed to take into account the possibility that the estimate $\tilde{s}_s[n]$ has been inverted.

The matrix elements in row $r$ and column $s$ of $[C_{ij}{}^q]$ are then eliminated and the above procedure is carried out on the remaining rows and columns of this matrix to determine $\varepsilon_i{}^q$ for the other actual sources.

The procedure in the previous paragraph is carried out for each Monte Carlo run.

There are two figures of merit that can be used to assess the performance of the processing methods:

*(1) Total RMS Estimation Error*

The RMS estimation errors at each time point, $n$, can be computed for each source by averaging $(\varepsilon_s{}^q[n])^2$ over all Monte Carlo runs and taking the square root:

$$RMS[n] = \sqrt{\frac{1}{Q} \sum_{q=1}^{Q} \left(\varepsilon_s{}^q[n]\right)^2} \qquad\qquad (30)$$

where $Q$ is the total number of Monte Carlo runs.



The total rms error over all data points can be calculated from

$$RMS_{tot} = \sqrt{\frac{1}{M} \sum_{n=1}^{M} \left[RMS[n]\right]^2} \tag{31}$$

*(2) Maximum RMS Estimation Error*

This is calculated from the maximum over $n$ of $RMS[n]$ for each component:

$$RMS_{\max} = \max_{n}\{RMS[n]\} \tag{32}$$

10 sets of 1000 Monte Carlo runs are carried out. $RMS_{tot}$ and $RMS_{max}$ are averaged over these 10 sets and the corresponding standard deviations are also calculated to assess the significance of these results.

To quantify the amount of noise that is added the following convention is used.

From Equation (1), the contribution of the $j^{th}$ source to the $i^{th}$ measurement is given by

$$p_{ij}[n] = A_{ij}\,s_j[n]$$

We can quantify the peak value (positive or negative) of the contribution of source $j$ to measurement $i$ by $P_{ij}{}^{MAX}$ where

$$P_{ij}{}^{MAX} = \max_{n}\left|p_{ij}[n]\right|$$

Now the performance of the algorithm will depend on how the noise standard deviation compares to the minimum value of $P_{ij}{}^{MAX}$, which is the smallest contribution of any source to the data mixtures:

$$R = \min_{i,j}\left\lfloor P_{ij}{}^{MAX}\right\rfloor \tag{33}$$

The noise standard deviation will be quoted as a percentage of $R$.

## 6.1    Example 1: Mixtures of Sparse Sources

It is found that for no noise, the MHC and Global methods are relatively insensitive to the choice of input parameters and the following are chosen for the sake of example:



$\alpha = 1$  (Equation (28)).

$v^{th} = 0.8$  (Equation (25)) for the MHC

$v^{th} = 0.4$  (Equation (25)) for the Global Method

In the absence of noise, it is found that the MHC, Global and Clusterwise PCA methods give almost perfect reconstruction of the original sources. The sources here are purely sparse and a good performance is to be expected for these algorithms which have been designed for such sources. However, it is found that there are significant estimation errors for both sources when applying Fast- ICA; in Figures 5(a) and 5(b) the estimated and actual sources 1 and 2 are shown; both the estimated and original sources have been normalised to unity rms values.

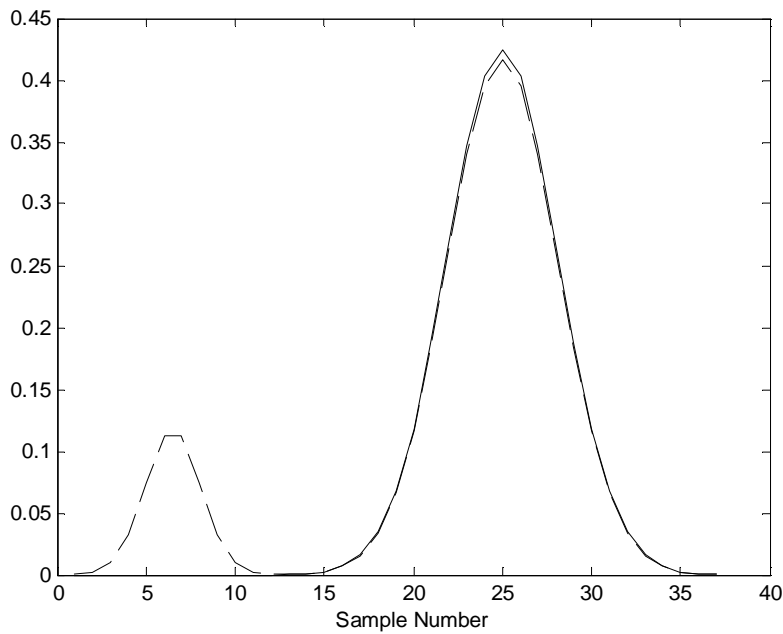

*Figure 5(a)- Comparison between the Original (Full Line) and FastICA Estimate (Dashed Line) for Source 1*



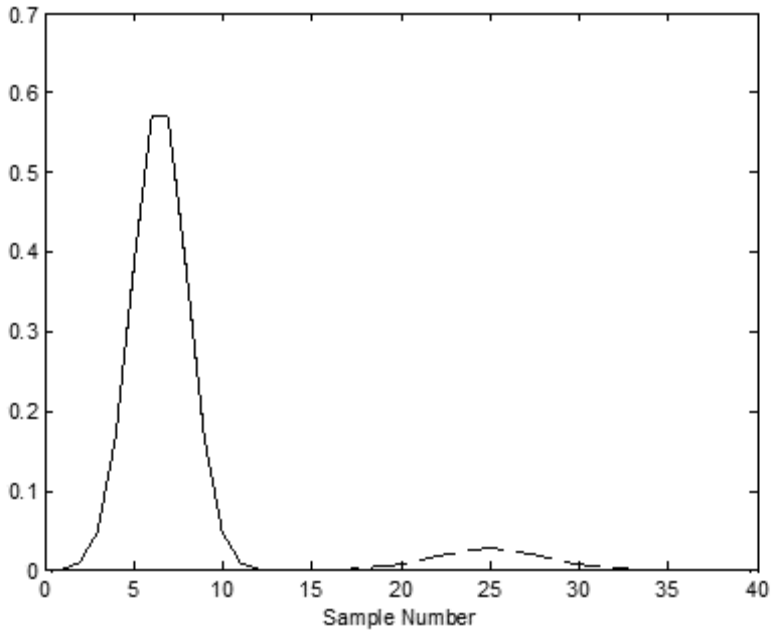

*Figure 5(b)- Comparison between the Actual (Full Line) and Fast-ICA Estimate (Dashed Line) for Source 2*

For Fast ICA, before applying PCA, the mean of the data is subtracted, which introduces correlations between the whitened sources; a consequence of this is that the estimate of one source being contaminated by a contribution from the other source. One can characterise the estimation error by quantifying the amplitude of the contaminating peak compared with the amplitude of the actual peak as a percentage; for Source 1 this percentage is 27% and the corresponding value for Source 2 is 4.7% . Simulations are next carried out for noise added to the measured data with sd = 0.005, which is 2.5% of $R$ defined in Equation (33). In this case, it is found that the performance of the Global method is sensitive to changes in the choice of $v^{th}$. If one tabulates the maximum of the RMS error, $RMS_{max}$ (32), averaged over 10 sets of 1000 Monte Carlo runs, as a function of $v^{th}$ then the results are as is as in Table 8. In this and subsequent tables, the numbers in brackets are the standard deviation of the rms errors computed over 10 sets of 1000 Monte Carlo runs.



| Method | Source 1 (x10$^3$) | Source 2 (x10$^3$) |
|---|---|---|
| Fast ICA | 112 (0.102) | 35.2 (0.391) |
| Clusterwise PCA | 6.765(0.102) | 27.6(0.441) |
| MHC vth = 0.7 | 11.7(0.379) | 27.0(0.216) |
| Global vth = 0.3 | 23.5(3.06) | 29.5(1.19) |
| Global vth = 0.35 | 9.74(2.55) | 26.8(0.241) |
| Global vth = 0.4 | 7.24(0.495) | 26.8(0.241) |

*Table 8 – Maximum of RMS Errors, noise sd* = 0.005

It can be seen that the Global method is sensitive to the parameter $v^{th}$ with a 68% reduction in maximum error for Source 1 as $v^{th}$ is increased from 0.3 to 0.4, with a corresponding reduction of 12% for Source 2. With $v^{th} = 0.4$, the Global Method has a comparable performance to Clusterwise PCA and a better performance than MHC and Fast ICA. For $v^{th} = 0.45$, it is found that the Global Method breaks down as clusters are unable to be formed.

Also looking at Table 8, it can be seen that, for the larger Source 1, the Clusterwise PCA and Global methods have the best performances, although one has to change $v^{th}$ for Global to an appropriate value. The MHC has a worse performance than these two methods because it only tries to find the best single "heading" to determine the dominant direction whilst the Global method operates on a cluster of headings and the Clusterwise PCA methods operate on the whole data. The Fast-ICA method still has the worst performance because of the subtraction of the mean but its performance is not significantly worse than with no noise indicating that it is more robust to noise than the other techniques. For the smaller Source 2, the four methods have comparable performances.

When the noise standard deviation has increased to 0.01, which is 5% of the maximum magnitude of the smaller source in the mixed data, then the results in Table 9 are obtained for the maximum rms errors for the various methods.



| Method | Source 1 (x10³) | Source 2 (x10³) |
|:---:|:---:|:---:|
| Fast ICA | 110.7(0.300) | 53.4(0.772) |
| Clusterwise PCA | 15.5(0.257) | 54.0(1.04) |
| MHC vth = 0.8 | 29.4(0.771) | 52.0(0.410) |
| Global vth = 0.3 | 34.5(1.47) | 52.7(1.09) |

*Table 9 – Maximum of RMS Errors, noise sd = 0.01*

All methods now give comparable performances for the estimation of Source 2. For Source 1, Clusterwise PCA yields the best results with MHC having the second best performance. The Global method is fairly insensitive to changes in the parameter $v^{th}$ and now yields worse results than the MHC method – averaging over noisy vectors does not yield better performance than the MHC. The Global $v^{th}$ method breaks down for $v^{th} \geq 0.4$ where this method is unable to form clusters.

Overall, taking into account both clean and noisy data, the Clusterwise PCA method has the best performance of the methods tested for this dataset. The Global method does have comparable performance to the Clusterwise method for clean data and also for noise standard deviation of 0.005 as long as an appropriate value for $v^{th}$ is chosen. For the largest noise standard deviation that is chosen, the Global method has an inferior performance when compared with the Clusterwise PCA and MHC methods.

## 6.2     Example 2: Mixture of Locally Sparse Sources

Sound signals, for example speech and music, can in many cases be considered as approximately sparse, so it would be of interest to see if the proposed method can separate out the individual sources from mixtures of sound signals. The sources will, in general, be overlapping and correlated. The set of sources that we will use are taken from [20]. Further details concerning these data can be found in References [21] and [22].



 The source files are taken from the development data and are wdrums_ src_1,wdrums_src_2 and wdrums_src_3 – these all represent music including drums.  The sampling frequency is 16 kHz.  The following randomly chosen mixing matrix is used:

$$A = \begin{pmatrix} -0.91935141 & 0.35931601 & -0.37350394 \\ -0.92906443 & 0.42466101 & 0.56348773 \\ 0.29719212 & 0.15967404 & -0.68715652 \end{pmatrix} \tag{34}$$

Data are taken between samples 19000 and 20000.  The correlation matrix of the sources, over this data segment, is given by

$$\begin{pmatrix} 1.0000 & 0.1937 & -0.0182 \\ 0.1937 & 1.0000 & 0.0077 \\ -0.0182 & 0.0077 & 1.0000 \end{pmatrix} \tag{35}$$

where the $(i,j)^{th}$ element is the correlation between sources $i$ and $j$.

The sources are not localised, as for Example 1, and so it is found that a better figure of merit to assess estimation errors is the total RMS estimation error (31).

It is found that the MHC method performs optimally for $v^{th} = 0.7$ in (25).  The RMS estimation errors for the Global method are calculated for $v^{th}$ varying from 0.3 to 0.6 in (25). In Table 10, the performance of the Global method is compared to the MHC, Clusterwise PCA and Fast ICA.

| Method | Source 1 (x10$^3$) | Source 2(x10$^3$) | Source 3(x10$^3$) |
|---|---|---|---|
| Fast ICA | 7.77 (0.113) | 12.9 (0.115) | 7.32(0.121) |
| Clusterwise PCA | 20.4(0.067) | 16.9(0.0553) | 25.3(0.0230) |
| MHC (vth = 0.7) | 7.19 | 1.85 | 3.52 |
| Global vth = 0.3 | 1.30 | 4.97 | 0.304 |
| Global vth = 0.4 | 0.868 | 5.38 | 0.526 |
| Global vth = 0.5 | 1.088 | 5.26 | 0.135 |
| Global vth = 0.6 | 1.96 | 4.57 | 0.466 |

*Table 10 – RMS Errors (clean signal)*



The Global method has the best performance of the methods tested when estimating Sources 1 and 3 but MHC is better at estimating Source 2. Fast ICA has a worse performance than these two methods and Clusterwise PCA has the worst performance. Now Clusterwise PCA is not working well, understandably, because the sources are not purely sparse. The subtraction of the mean prior to applying the Fast ICA is causing additional correlations to appear between sources; in addition, even without this subtraction, it can be seen from the matrix (35) that there is significant correlation between sources 1 and 2.

In this example, we have a mixture of signals where there are no segments where one signal exists on its own – hence the sources are not purely sparse. However there are segments where each source is dominant. When applying the MHC or Global methods, the estimate of heading for one source will consists of contaminations from other sources. The Global method averages over these contaminations, treating them as noise, so, in theory, should be better than local MHC, which chooses one heading only; this is the case for Sources 1 and 3, but not for Source 2 which has small amplitude so averaging here does not help because of the relatively large values of the contaminations.

In Figure 6, the normalised sources and the errors (29) in the estimates from the Global Method ($v^{th} = 0.5$in (25)) are plotted together as a function of sample number. It can be seen that the errors for Sources 1 and 3 are almost imperceptible, whilst the errors for Source 2 are more significant.



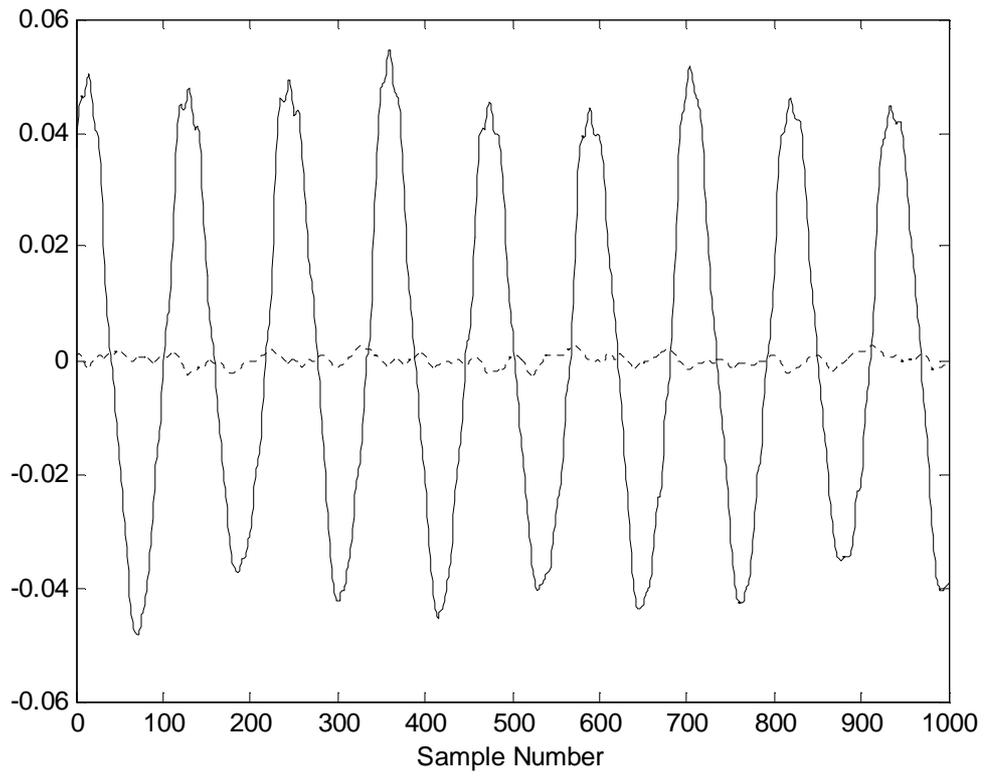

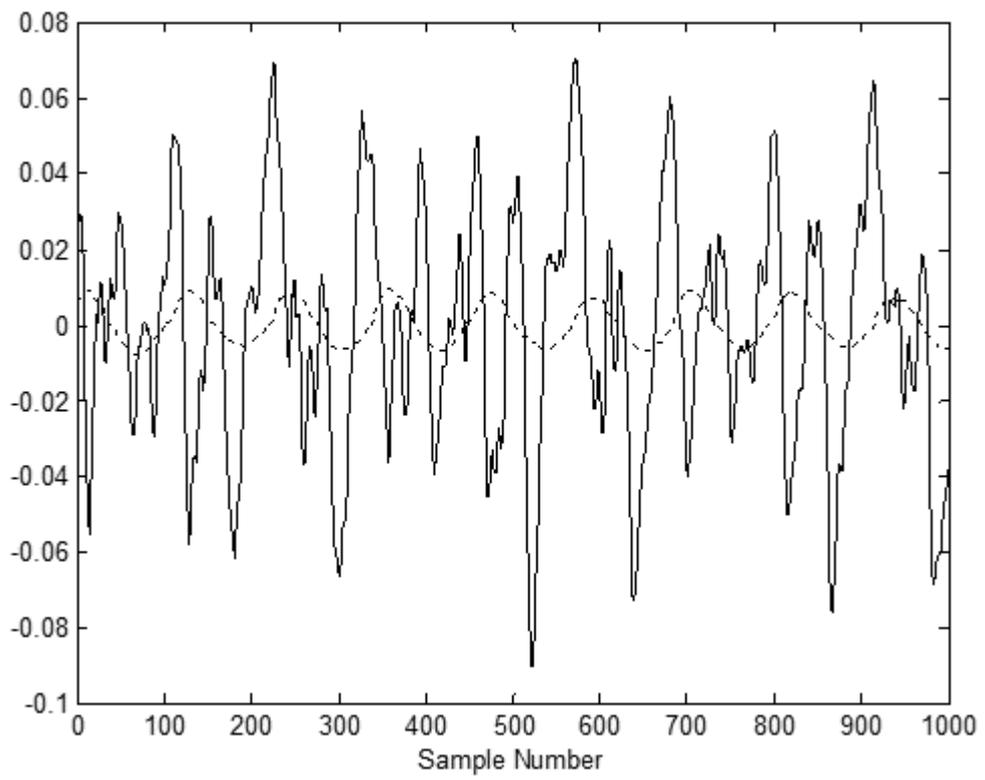



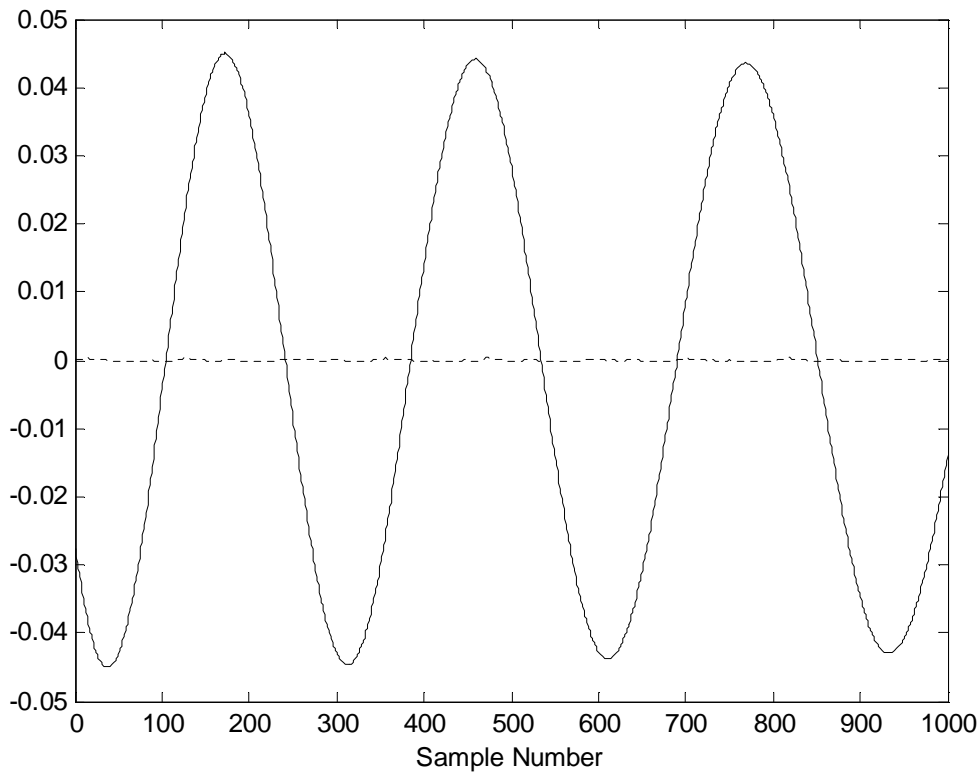

*Figure 6 – Comparison between actual source (full line) and error in estimated source (dashed line) for each source: Top: Source 1, Middle: Source 2: Bottom: Source 3. All actual and estimated sources have been normalise to an rms value of 1.*

In Table 11, the results are shown for the case where noise with standard deviation of 0.0005 is added to the mixtures corresponding to 20% of $R$ given in Equation (33).

| Method | Source 1 (x$10^3$) | Source 2(x$10^3$) | Source 3(x$10^3$) |
|---|---|---|---|
| Fast ICA | 7.52(0.267) | 17.4(0.201) | 6.82(0.214) |
| Clusterwise PCA | 22.7(0.116) | 23.1(0.281) | 23.4(0.065) |
| MHC (vth = 0.7) | 7.83(0.094) | 13.6(0.012) | 7.10(0.095) |
| Global vth = 0.3 | 4.58(0.014) | 13.6(0.009) | 3.65(0.016) |
| Global vth = 0.4 | 4.32(0.015) | 13.6(0.008) | 3.37(0.017) |
| Global vth = 0.5 | 4.08(0.018) | 13.6(0.013) | 3.11(0.019) |

*Table 11 – RMS Errors (Noise sd = 0.0005)*

Comparing Tables 10 and 11, it can be seen that the Global and MHC methods are more sensitive to noise than the Clusterwise PCA and Fast ICA methods. The reason for this is that



the Fast ICA and Clusterwise PCA methods process the whole data.  The localised MHC processes one heading at a time and is hence not using the whole data to process the signal – this makes it more sensitive to noise than Clusterwise PCA and Fast ICA.  The Global method  clusters part of the data so has better performance in noise than Local MHC for Sources 1 and 3 and is comparable with MHC for Source 2 consistent with the case when no noise is added.

Finally, it should be noted that, when the MHC and Global methods are applied to the clean signal, the sources are estimated in the order 2, 1, 3.  Source 2 has highest frequency components, followed by Source 1 and then Source 3. This can be explained as both the MHC and Global methods operate on the "headings" which are taken from differences between adjacent data points. This differencing operation is equivalent to applying a high-pass filter to the data which will accentuate the highest frequency components.  This may also explain the relative sensitivity of the MHC and Global methods to noise.

## 6.3     Example 3: 8-Lead Thoracic and Abdominal ECG Data from Expectant Mother

In order to compare the performances of the FastICA and Global methods, data taken from the Daisy Database [23] will be analysed. This data consists of ECG signals taken from an expectant mother. The data consists of 8 leads, 1 to 5 being abdominal and 6 to 8 thoracic. The first 1000 samples of the data are chosen; there is some uncertainty about the sampling frequency, also pointed out in [24], but it is probably 250 Hz.

Data from a typical abdominal lead is shown in Figure 7.



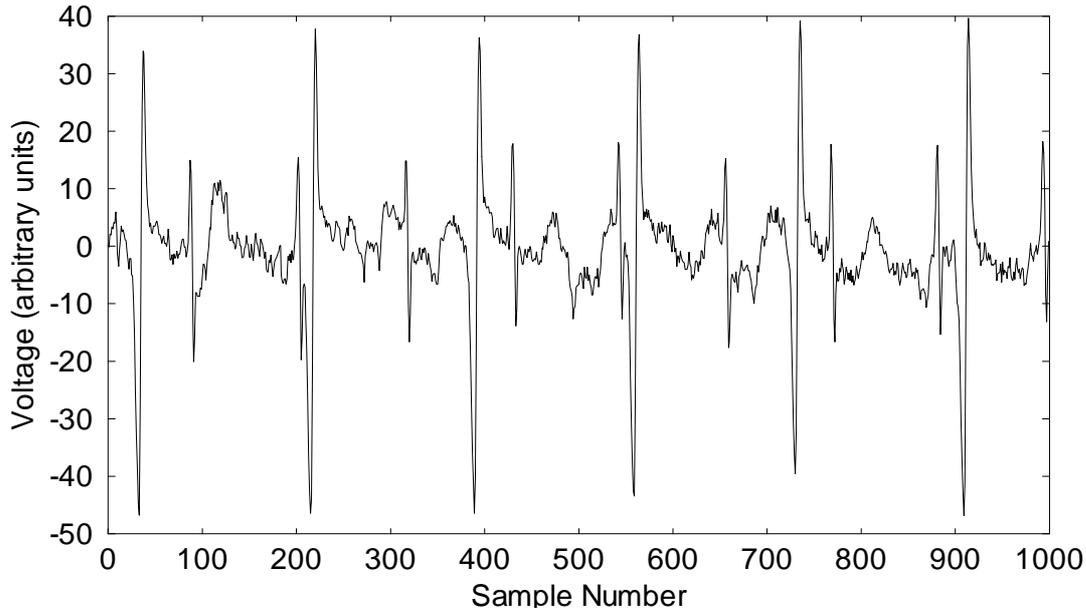

*Figure 7 – Lead 1*

These data represent a more demanding test for the proposed technique: for multi-lead ECG data, the morphology of each source can change with time along with the amplitude. In addition, there is noise present which will affect the performance of the method. In previous work [15,16], the localised MHC method gave comparable maternal and fetal outputs compared with the ICA method. The component most resembling the fetal signal is chosen by visual inspection. It is found that, when applying the Global method, the quality of the extracted fetal signal is sensitive to the choice of $v^{th}$. In Figures 8(a) to 8(c), the component that looks most fetal is displayed for choices of threshold parameter $v^{th} = 0.1$, $0.2$ and $0.3$ in (25). It can be seen that for $v^{th} = 0.1$ and $0.3$, a good quality fetal component is extracted. However for $v^{th} = 0.2$ the quality of the fetal component is much worse.

In Figure 8(d), the corresponding component extracted using FastICA is shown where the deflation method is used along with Gaussian non-linearity. A good quality fetal component is extracted comparable to those extracted by the Global method for $v^{th} = 0.1$ and $0.3$. Now, when applying Fast ICA, one has the option to apply the deflation method or the symmetric



method; in addition, one has the choice of the following non-linearities: Gaussian, pow3, tanh and skew [7,8]. It is found that the Fast-ICA consistently extracts fetal components with similar qualities to Figures 8(a) and 8(c) regardless of the combination of method/non-linearity that is being used and also independently of the random choice of initial weights. Hence, although it is possible to extract fetal components using the Global method that are of similar quality to those extracted using Fast-ICA, the latter method is more robust to input parameters than the Global Method.

**(a)**

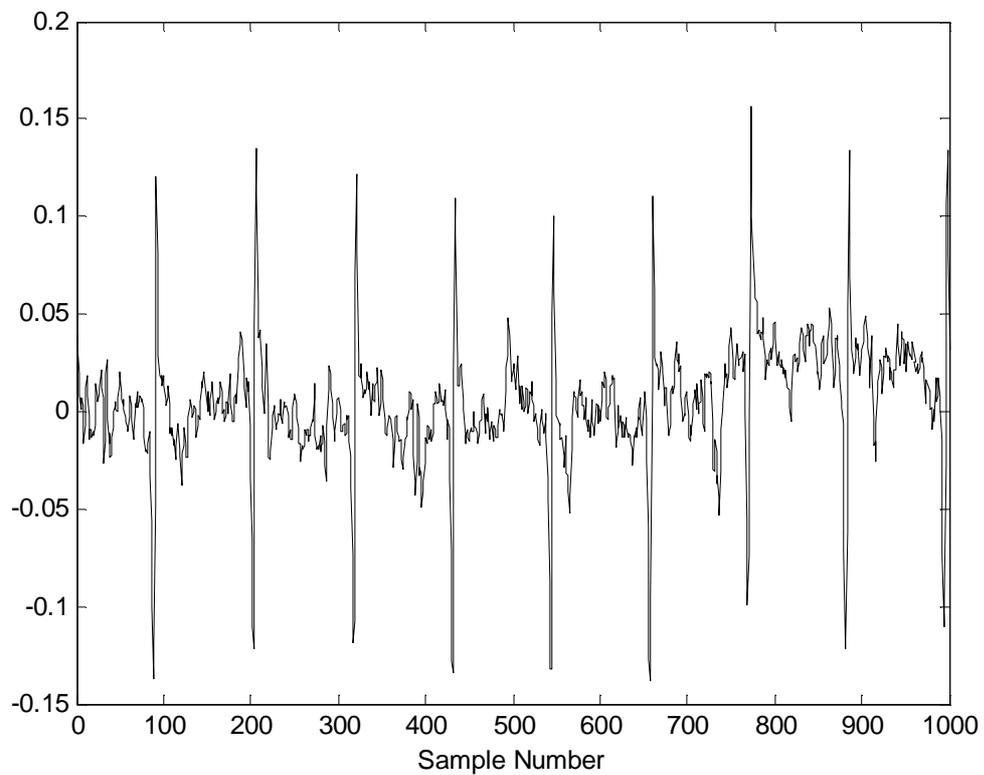



**(b)**

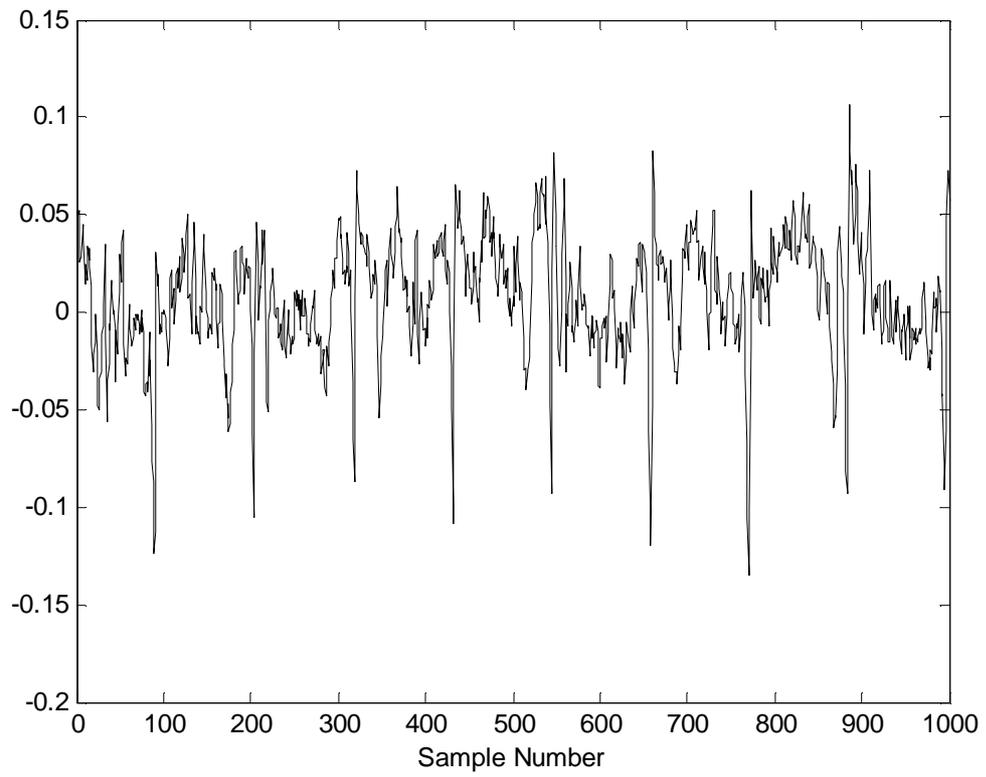

**(c)**

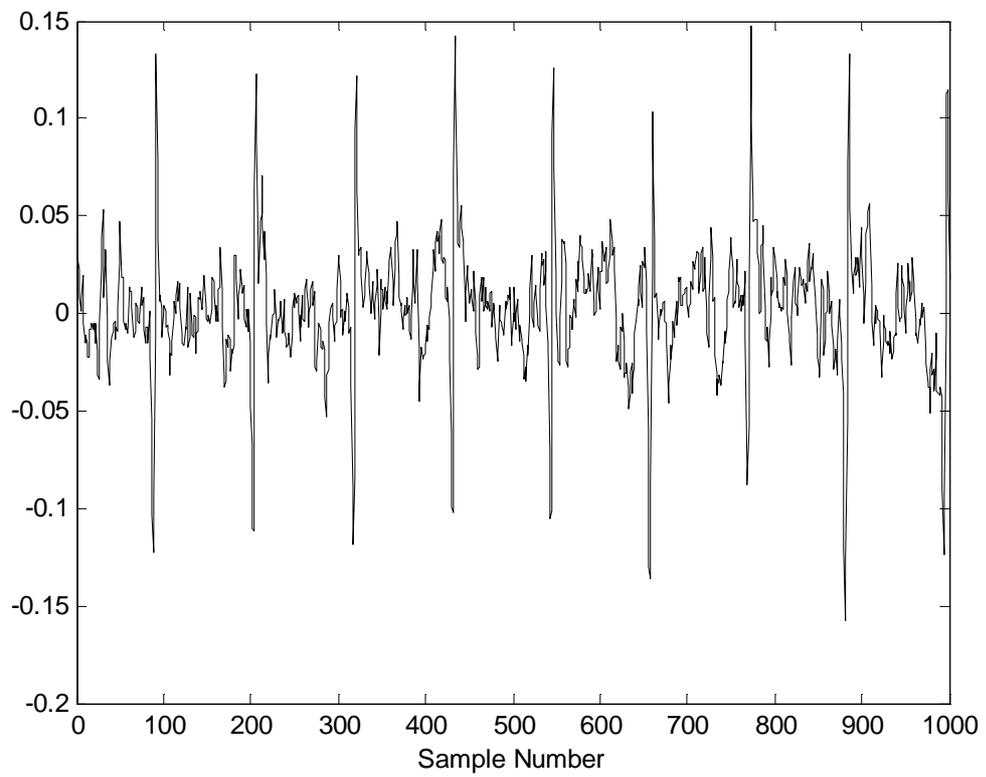



**(d)**

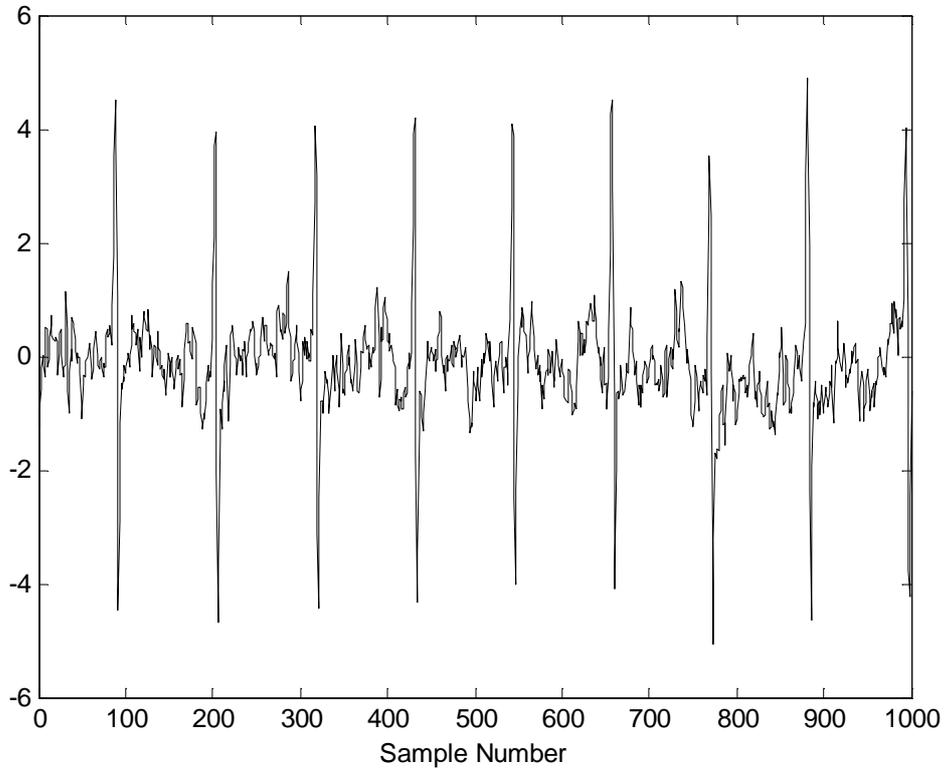

*Figure 8- "Best" fetal estimation (a) Global vth =0.1, (b) Global vth = 0.2, (c) Global vth = 0.3, (d) Fast ICA*

## 7    Discussion

The novel method presented in this paper is one of a family of methods that have been specifically developed to separate sparse sources from mixtures of such signals [9-17]. In the particular approach that is adopted in this paper, (and in previous references), one first determines the segments of data where one source dominates and then one uses that information to extract the sources one-by-one. The potential advantage of this approach over more general approaches to BSS is that one is concentrating on data where one can obtain most information on the sources which should hopefully lead to better extraction.



The approach used in this paper has been successful in extracting sources for the examples that are studied in this paper. In some cases, better extraction than Fast-ICA has been achieved. However, the two main drawbacks of this method are:

(1)      Sensitivity to input parameter $v^{th}$ (25)

(2)      Sensitivity to noise.

It is observed with Example 2 (Section 6.2) that, when using Global and MHC methods, the more rapidly varying components are detected first. This is due to the headings being calculated from differences of data; hence one is applying a high pass filter. The advantage of this approach for Signal 2 is that the highest frequency source is not sparse, but because the other sources were effectively attenuated, the filtered version becomes sparser and is hence easier to detect. The dominance of each frequency component in turn means that each component is detected as against all the other components. The disadvantage of the high pass filtering action is that high frequency components of noise are amplified leading to relatively poor performance in noise compared to the other methods.

The performances of the Fast-ICA and Clusterwise PCA methods are much less sensitive to input parameters. This means that, when implemented on-line, these two methods would have a more reliable performance than the Global and MHC methods because of the uncertainty in the best choice of input parameter to use for a specific signal and signal to noise ratio.

## 8      Conclusions

A blind source separation method has been developed for mixtures of sparse sources. The method involves identifying data points that are dominated by one source and to use this information to estimate that source by estimating a dominant direction in phase space. Principal directions in the phase plot are estimated using a simple clustering technique based on sorting the heading components in ascending order of magnitude. This estimated source is



then subtracted from the data and the process repeated to estimate the other sources. This method is an extension of the work in [15] where the headings are chosen at the point where there is least change between two adjacent headings.

The method has been evaluated on simulated data where the sources are sparse, simulated data where they are semi-sparse and experimental fetal ECG monitoring data. The proposed method, called the Global method, is more robust to noise than the MHC method [15] and can lead to comparable and sometimes better estimates than the standard ClusterwisePCA and Fast-ICA methods. However the proposed method (like the MHC) is sensitive to the choice of input parameter $v^{th}$ (25) and so may be best implemented off-line along where one has the time to perform several analyses of data with different input parameters. The Global method, whilst more robust to noise than the MHC, can be more sensitive to noise than Fast-ICA and ClusterwisePCA; this is because the Global method processes only part of the data whilst the Fast-ICA and ClusterwisePCA process the whole data.

The Global method has been developed assuming that the sources are uncorrelated so that, at each iteration, it is possible to extract the sources one-by-one. In Example 2 in Section 6.2, where the sources are weakly correlated, the Global Method performed relatively well because when estimating the heading for one source, the other sources contributed what was effectively noise to the estimate of that heading which can then be averaged over using Equation (23). However, for strongly correlated sources the method will break down. The MHC method has been extended in [16] to the case of correlated sources, where it is assumed that all sources are sparse. It would be of interest to see whether the Global method can be extended in a similar way to deal with correlations between the underlying sources.




**REFERENCES**

[1]    Ren, J.Y., Chang, C.Q., Fung, P.C.W., Shen, J.G., and Chan, F.H.Y.:
       'Free Radical EPR Spectroscopy Analysis using BSS', *Journal of Magnetic
       Resonance*, 2004, vol.166, (1), pp 82-91

[2]    Yin P, Sun Y and Xin J, 'A Geometric Blind Source Separation Method Based on
       Facet Component Analysis', *Signal, Image and Video Processing*, 2016, vol. 10, pp
       19-28

[3]    Varanini M, Tartarisco G, Billeci L, Macerata A, Pioggia G and R Balocchi R, 'An
       efficient unsupervised fetal QRS complex detection from abdominal maternal ECG',
       *Physiological Measurement*, vol. 35, 2014, pp 1607–1619

[4]    Dawy, Z., Sarkis, M., Hagenauer, J., and Mueller, J.: 'A Novel Gene Mapping
       Algorithm based on Independent Component Analysis', *IEEE International
       Conference on Acoustics, Speech, and Signal Processing (ICASSP) 2005 Conference*,
       March 18-23, 2005, Volume V, pp. 381-384

[5]    Shlens J, 'A Tutorial on Principal Component Analysis'
       http://arxiv.org/abs/1404.1100v1 (2014) - Last accessed March 11[th] 2016

[6]    Zarzoso, V., and Nandi,A.K.: 'Blind Separation of Independent Sources for Virtually
       Any Source Probability Density Function', *IEEE Transactions on Signal Processing*,
       1999, vol. 47, (9), pp 2419-2432

[7]    Hyvärinen A. 2013 'Independent Component Analysis: Recent Advances. Phil Trans
       R Soc A 371: 20110534. http://dx.doi.org/10.1098/rsta.2011.0534

[8]    Hyvärinen A and Oja E, 'Independent component analysis: algorithms and
       applications', *Neural Networks*, 2000, vol. 13, no. 4-5, pp. 411–430





[9] Babaie Zadeh, M., Jutten, C., and Mansour, A., 'Sparse ICA via Cluster-wise PCA', *Neurocomputing* ,2006, vol. 69, (13-15),  pp.1458-1466

[10] Chang C, Fung P C W and Hung Y S , 'On a Sparse Component Analysis Approach to BSS' , *6th International Conference on Independent Component Analysis and Blind Source Separation  (ICA 2006),* Charleston, South Carolina, USA, March 2006, pp 765-772

[11] O'Grady, P.D., Pearlmutter, B.A., Rickard, S.T. 'Survey of Sparse and non-Sparse Methods in Source Separation', *International Journal of Imaging Systems and Technology*, 2005, vol. 15, (1), pp. 18-33

[12] Theis, F., Jung, A., Puntonet, C., and Lang, E.W. 'Linear Geometric ICA: Fundamentals and Algorithms', 2003, *Neural Computation*, 15, (2), pp. 419-439

[13] Georgiev, P., Theis, F., and Cichocki, A.: 'Sparse Component Analysis and Blind Source Separation of Underdetermined Mixtures', 2005, *IEEE Transactions on Neural Networks,* 16, (4), pp. 992-996

[14] Davies, M., and Mitianoudis, N., 'Simple Mixture Model for Sparse Overcomplete ICA', (2004), *IEE Proceedings on Vision, Image and Signal Processing*,  2004, 151,(1), pp. 35-43

[15] Woolfson, M.S., Bigan , C., Crowe, J.A. and Hayes-Gill, B.R. 'Method to separate sparse components from signal mixtures', 2008, *Digital  Signal Processing*, 18, (6), pp. 985-1012

[16] Woolfson M. S,  Bigan C, Crowe J. A., and Hayes-Gill B. R., "Extraction of Correlated Sparse Sources from Signal Mixtures," *ISRN Signal Processing*, vol. 2013, Article ID 218651, 17 pages, 2013. doi:10.1155/2013/218651





[17]    Sun, Y., Ridge C., del Rio F., Shaka A.J. and Xin, J.: ' Postprocessing and sparse

blind source separation of positive and partially overlapped data', 2011, *Signal*

*Processing*, vol. 91(8), pp. 1838-1851

[18]    Davila C E  and Mobin M S,

'Weighted Averaging of Evoked Potentials', 1992, *IEEE Transactions on Biomedical*

*Engineering*, vol. 39, no. 4, pp 338-345

[19]    http://research.ics.aalto.fi/ica/fastica/ - last accessed March 11[th] 2016

[20]    'Stereo Audio Source Separation Evaluation Campaign', SASSEC07,

http://www.irisa.fr/metiss/SASSEC07/

Last accessed March 11[th] 2016

[21]    Vincent E, Araki S,  Theis F J, Nolte G, Bofill P, Sawada H, Ozerov A,

Gowreesunker B V, Lutter D and Duong N Q K, ' The Signal Separation Evaluation

Campaign (2007-2010): Achievements and remaining challenges', 2012, *Signal*

*Processing*, vol. 92, pp. 1928-1936

[22]    Vincent E, Sawada H, Bofill P, Makino S and Rosca J P, 'First stereo audio source

separation evaluation campaign: data, algorithms and results', in Proc. Int. Conf. on

Independent Component Analysis and Signal Separation, pp. 552-559, 2007.

[23]    Database for the identification of systems (DaISy),

http://www.esat.kuleuven.ac.be/sista/daisy

Last accessed March 11[th] 2016

[24]    Zarzoso V. and Nandi A.K., "Non-invasive Fetal Electrocardiogram Extraction: Blind

Source Separation versus Adaptive Noise Cancellation", 2001, *IEEE Transactions on*

*Biomedical Engineering*, vol. 48, No 1,12-18